\documentclass[10pt,a4paper]{article}
\usepackage[utf8]{inputenc}
\usepackage[T1]{fontenc}
\usepackage{times}
\usepackage{amsmath,amssymb,amsfonts}
\usepackage{graphicx}
\usepackage{caption,subcaption}
\usepackage{multirow}
\usepackage{multicol}
\usepackage{authblk}
\usepackage{multicol}
\usepackage{xcolor}
\usepackage{multicol}
\usepackage[margin=2cm]{geometry}
\usepackage[singlespacing]{setspace}
\usepackage[sort, numbers]{natbib}
\usepackage{hyperref}
\usepackage[mathlines]{lineno}
\begin{document}

\title{KHIFC-user friendly program for studying Heavy ion fusion barrier characteristics}
\author[$1\dagger$]{H.C. Manjunatha}
\author[2]{P.S.Damodara gupta}
\author[$3\ddagger$]{N.Sowmya}
\author[4]{K.N.Sridhar}
\affil[1]{Department of Physics, Govt. First Grade College, Devanahalli-562110, Karnataka, India}
\affil[2]{Department of Physics, Rajah Serfoji Government College, Thanjavur-613005, Affiliated to Bharathidasan University, Tiruchirappalli-TamilNadu}
\affil[3]{Department of Physics, Govt. First Grade College, Chikkaballapur-562101, Karnataka, India}
\affil[4]{Department of Physics, Govt. First Grade College, Malur-563130, Karnataka, India}
\affil[*]{Corresponding author : $^\dagger$manjunathhc@rediffmail.com, $^\ddagger$sowmyaprakash8@gmail.com}

\date{\today}
\maketitle
\begin{abstract}
Abstract
We have developed an application for studying heavy ion fusion barrier characteristics such as such as fusion barrier heights ($V_B$), positions ($R_B$) and curvature of the inverted parabola ($\hbar\omega$). We call this application as KHIFC (Kolar Heavy Ion fusion barrier Characteristics). This software application hosted in the domain "https://systematics-of-heavy-ion-fusion.vercel.app/". This KHIFC produces fusion cross-sections with the simple input of projectile, target and center of mass energy. The values produced by the  KHIFC is validated with experiments. Efficient tools like KHIFC are essential for researchers in nuclear physics, particularly when dealing with complex systems such as actinide and superheavy nuclei. By providing quick calculations and insights, it can significantly aid in experiment planning and theoretical investigations.
\end{abstract}


\section{PROGRAM SUMMARY}
Program title: KHIFC\\
CPC Library link to program files:https://data.mendeley.com/datasets/48gwwtfc7z/2\\
Licensing provisions: Creative Commons by 4.0 (CC by 4.0) \\
Programming language: JavaScript, HTML, CSS\\
Nature of problem: Evaluating the characteristics of the fusion barrier involves a multi-step and complex process that requires careful calculations.\\
Solution method: By using React with JavaScript, the developed software accepts inputs and evaluates the characteristics of the fusion barrier based on the given input. It also generates graphs to visualize the fusion cross-section data.

\section{Introduction}\label{introduction}
\indent The synthesis of heavy or superheavy nuclei is crucial and is motivated by the theoretical expectation of an "island of stability." These nuclei are the subject of much experimental research, mostly via fusion-evaporation processes \cite{hofmann2000g,oganessian2007heaviest,oganessian2007synthesis}. Elements 113 to 118 are among the discoveries made possible by laboratories worldwide. Despite the difficulties in creating and researching these elusive and transient nuclei, the search for superheavy elements is still important and might impact nuclear physics and chemistry.
The cold fusion reactions utilize lead and bismuth as targets for Z=113 \cite{hofmann1995production,morita2004status,munzenberg1984evidence,morita2012new}, while hot fusion employs $^{48}Ca$ projectiles with actinide targets for Z$\ge$114 \cite{oganessian2004measurements,oganessian2012new,oganessian2012production}. Identifying superheavy elements relies on confirming consistent decay chains of the produced compound nucleus. \\
\indent After the synthesis of superheavy element Z=118, many experiments were carried out for the synthesis of superheavy element Z=119 and 120 \cite{hessberger2001decay, hamilton2013search}. Additionally, many theoretical models such as dinuclear system model (DNS) \cite{adamian2003characteristics,zhu2015production}, Advanced statistical model (ASM) \cite{manjunatha2020entrance,manjunatha2022role}, dynamical cluster-decay model \cite{chopra2015determination}, dynamical approach based
on Langevin equations \cite{shen2002two}, Time-dependent Hartree-Fock model \cite{sekizawa2019time}, improved quantum molecular dynamics model \cite{wang2004further}, fusion-by-diffusion model \cite{siwek2012predictions}, and many were used in the prediction of evaporation residue cross-sections.\\
\indent The fusion is a complex process and the cross-sections can vary depending on the specific reaction and conditions involved. To understand the entire fusion process in heavy-ion collisions, \citet{iwamoto1996collisions} considered the influence of the deformation of both projectile-target nuclei. A comprehensive literature survey reveals theoretical studies focusing on fusion barriers in low and medium atomic number nuclei \cite{vaz1981fusion,puri1992fusion,arora2000analytical,sahu1999asymmetric,moustabchir2001analytic,stefanini1986heavy}. In addition many codes such as CASCADE \cite{puhlhofer1977interpretation}, HIVAP \cite{reisdorf1981analysis,reisdorf1985fusability,reisdorf1981analysis}, CCFUL \cite{hagino1999program}, PACE \cite{gavron1980statistical}, and
POTFUS + GEMINI++ \cite{mancusi2009comparison} has been used to predict fusion cross-sections. 
Codes provide estimates of fusion cross-sections as a function of energy and other relevant parameters. The importance of these codes in predicting fusion cross-sections lies in their ability to guide experimental efforts, optimize conditions for fusion reactions, and advance our fundamental understanding of nuclear physics.\\
\indent Quantum mechanical effects, such as tunneling, play a significant role in nuclear fusion. Tunneling allows nuclei to overcome the Coulomb barrier even when their kinetic energy is insufficient to do so classically. These fusion cross-sections depend sensitively on the energy of the colliding nuclei. Predicting cross-sections across a range of energies requires sophisticated models that account for energy dependence. Hence, this paper aims to address a notable gap in existing literature where references to nuclear codes often lack comprehensive information about the intricacies of the calculations, creating challenges for new users in grasping the inner workings of these codes and discerning their range of applicability. So, the primary objective of this paper is to offer an exhaustive and comprehensive account of each step essential for executing a fusion cross-sections in the actinide and superheavy region, thereby facilitating a thorough understanding for researchers and practitioners entering this domain. \\
\section{Theoretical Framework}\label{theory}
\indent In nuclear fusion, the fusion barrier characteristics such as fusion barrier height, barrier position, and the concept of an inverted parabola, are crucial in understanding the dynamics of fusion reactions. The fusion barrier height denotes the energy required for atomic nuclei to overcome electrostatic repulsion during fusion. The barrier position represents the distance where the barrier is maximal, influenced by nuclear forces. An inverted parabola in nuclear fusion depicts the potential energy between colliding atomic nuclei. At close distances, attractive forces prevail, while repulsive forces dominate as nuclei separate. This curve illustrates the energy barrier for fusion reactions, which is crucial for predicting fusion dynamics and optimizing experimental conditions. Further, theoretical formalism for fusion barrier characteristics Actinides \cite{manjunatha2018fusion} and superheavy nuclei \cite{manjunatha2018pocket} were adopted from previous work. The constructed theoretical formalism for active region involves 7205 projectile-target combinations within specified ranges for atomic and mass numbers ($3 < Z_1 < 51, 45 < Z_2 < 100$ and $6 < A_1 < 127, 101 < A_2 < 260$). Where as in case of superheavy nuclei ($104 < Z < 130$), The used theoretical formalism involves 14054 projectile target combinations with atomic and mass number ranges $6 < Z_1 < 58, 52 < Z_2 < 98$ and $12< A_1 < 142, 120 < A_2 < 252$. The adopted Fusion Characteristics such as fusion barrier heights ($V_B$), positions ($R_B$), curvature of the inverted parabola ($\hbar\omega$) are
\begin{align}
    V_B=1.866+1.435\times\left[\frac{Z_1Z_2}{R_B}\left(1-\frac{1}{R_B}\right)\right]
\end{align}
\begin{align}    \hbar\omega=1.46\times10^{-7}\left(\frac{Z_1Z_2}{A_1^{1/3}+A_2^{1/3}}\right)^3\\\nonumber-9.4\times10^{-5}\left(\frac{Z_1Z_2}{A_1^{1/3}+A_2^{1/3}}\right)^2\\\nonumber+1.02\times10^{-2}\left(\frac{Z_1Z_2}{A_1^{1/3}+A_2^{1/3}}\right)-4.02
\end{align}
Similarly, The adopted Fusion Characteristics for superheavy are
\begin{align}    S_B=-1.236\times10^{-7}\left(\frac{Z_1Z_2}{A_1^{1/3}+A_2^{1/3}}\right)^3\\\nonumber+7.774\times10^{-5}\left(\frac{Z_1Z_2}{A_1^{1/3}+A_2^{1/3}}\right)^2\\\nonumber-2.324\times10^{-2}\left(\frac{Z_1Z_2}{A_1^{1/3}+A_2^{1/3}}\right)+3.759
\end{align}
\begin{align}
V_B=1.4057\times\left[\frac{Z_1Z_2}{R_B}\left(1-\frac{1}{R_B}\right)\right]+5.4746
\end{align}
\begin{align}
\hbar\omega=-3.34\times10^{-7}\left(\frac{Z_1Z_2}{A_1^{1/3}+A_2^{1/3}}\right)^3\\\nonumber+1.39\times10^{-4}\left(\frac{Z_1Z_2}{A_1^{1/3}+A_2^{1/3}}\right)^2\\\nonumber-2.37\times10^{-2}\left(\frac{Z_1Z_2}{A_1^{1/3}+A_2^{1/3}}\right)+5.67
\end{align}
We have these fusion barrier characteristics such as fusion barrier heights ($V_B$), positions ($R_B$) and curvature of the inverted parabola ($\hbar\omega$) in wong formalism \cite{wong1973interaction}to obtain the fusion cross section;
\begin{align}
\sigma_{fus}=\frac{(\hbar\omega)^{}{(R_B^2)^{}}}{0.1E_{cm}} ln\bigg(1+exp\bigg[\frac{2\pi}{(\hbar\omega)^{}}(E_{cm}-V_B)^{}\bigg]\bigg)
\end{align}
\section{Results and discussion}\label{RD}
\indent The user-friendly, simple computer program KHIFC (Kolar Heavy ion fusion barrier characteristics) is developed for the study of heavy ion fusion systematics based on the theoretical formalism explained in section II. This program produces the fusion barrier characteristics such as fusion barrier height ($V_B$ in MeV), fusion barrier radius ($R_B$ in fm), inverted parabola ($\hbar \omega$ in MeV), fission barrier of the compound nucleus (Bf in MeV), and fusion cross-section.  This program will produce the fusion cross section corresponding to input $E_{cm}$ of a wide 100 MeV range. It also gives entrance channel parameters such as the Coulomb interaction parameter, mean fissility, mass asymmetry, charge asymmetry, and isospin number.  \\
\begin{table}[!h]
\caption{Comparison of fusion barriers (MeV) produced by KHIFC with that of the experiments}
\label{01}
\centering
\begin{tabular}{|l|l|l|}
\hline
\multicolumn{1}{|c|}{\textbf{Reaction}} & \multicolumn{1}{c|}{\textbf{Expt.}} & \multicolumn{1}{c|}{\textbf{KHIFC}} \\ \hline
$^{16}O+^{208}Pb\to^{224}Th$ & 74.55 \cite{morton1999coupled} & 75.362 \\ \hline
$^{19}F+^{208}Pb\to^{227}Pa$ & 83 \cite{hinde1999limiting} & 83.763 \\ \hline
$^{28}Si+^{208}Pb\to^{236}Cm$ & 128.1 \cite{hinde1995competition} & 128.408 \\ \hline
$^{48}Ca+^{208}Pb\to^{256}No$ & 173.40$\pm$0.1 \cite{banerjee2019mechanisms} & 176.224 \\ \hline
$^{54}Cr+^{208}Pb\to^{262}Sg$ & 207.30$\pm$0.3 \cite{banerjee2019mechanisms} & 210.786 \\ \hline
$^{54}Cr+^{208}Pb\to^{262}Sg$ & 205.8 \cite{mitsuoka2007barrier} & 210.786 \\ \hline
$^{56}Fe+^{208}Pb\to^{264}Hs$ & 223.00 \cite{dutt2010systematic, mitsuoka2007barrier} & 228.672 \\ \hline
$^{32}S+^{232}Th\to^{264}Ds$ & 155.73 \cite{dutt2010systematic} & 158.036 \\ \hline
$^{64}Ni+^{208}Pb\to^{272}Ds$ & 236.00 \cite{dutt2010systematic} & 243.25 \\ \hline
$^{40}Ar+^{238}U\to^{278}Ds$ & 171.0 \cite{bass1973threshold} & 177.352 \\ \hline
$^{58}Ni+^{208}Pb\to^{266}Ds$ & 236.0 \cite{mitsuoka2007barrier} & 246.725 \\ \hline
$^{70}Zn+^{208}Pb\to^{278}Cn$ & 250.00 \cite{dutt2010systematic} & 259.018 \\ \hline
$^{70}Zn+^{208}Pb\to^{278}Cn$ & 250.6 \cite{mitsuoka2007barrier} & 259.018 \\ \hline
$^{86}Kr+^{208}Pb\to^{294}Og$ & 299.00 \cite{dutt2010systematic} & 308.386 \\ \hline
$^{86}Kr+^{208}Pb\to^{294}Og$ & 303.3 \cite{ntshangase2007barrier} & 308.386 \\ \hline
$^{84}Kr+^{232}Th\to^{316}126$ & 332.0 \cite{bass1973threshold} & 337.509 \\ \hline
$^{84}Kr+^{238}U\to^{322}128$ & 333.0 \cite{moretto1972shellmodel} & 344.623 \\ \hline
\end{tabular}
\end{table}
\begin{figure}[]
    \centering
    \includegraphics[width=\linewidth]{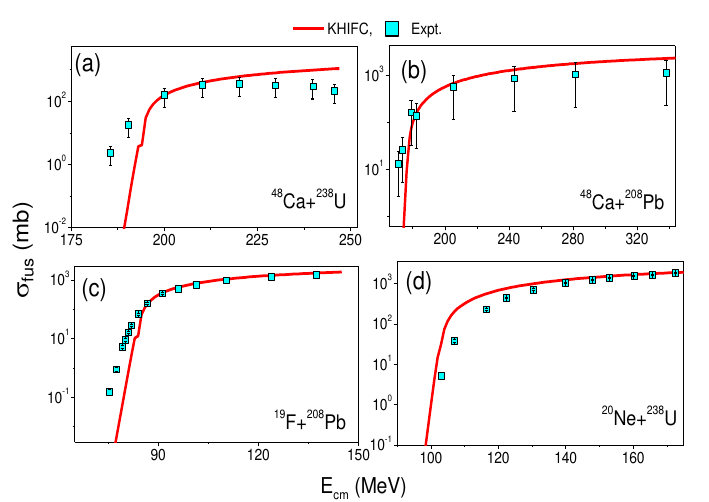}
    \caption{Comparison of fusion cross-section produced by KHIFC with that of the experiments.}
    \label{01}
\end{figure}
\indent The values produced by the present program is validated with the experiments. The fusion barriers and fusion cross sections produced by the KHIFC is compared with the experiments. table \ref{01} shows comparison of the fusion barriers produced by the KHIFC with that of the experiments. From this comparison it is observed that the fusion barriers produced by KHIFC is having the percentage of deviation less than 2.5\%. This program can be used to produce the fusion barriers of the compound nucleus up to Z=128. Furthermore, to validate the fusion-crosssections produced by KHIFC, fusion cross-sections  are compared with the experiments and this comparison is shown in fig. \ref{01}. For instance, we have selected the fusion reactions such as  $^{48}$Ca+$^{238}$U, $^{48}$Ca+$^{208}$Pb, $^{64}$Ni+$^{208}$Pb, and $^{40}$Ar+$^{238}$U and computed the fusion cross-sections using KHIFC. The comparison of experimental fusion cross-sections with that of the KHIFC is validates the fusion cross-sections. From this comparison it is observed that there is a deviation of KHIFC and experiments below the barrier may be due to the fact that deformations effect and angular moment effects are not considered. \\
\subsection{How to use the program}
\begin{figure}[]
    \centering
    \includegraphics[width=1\linewidth]{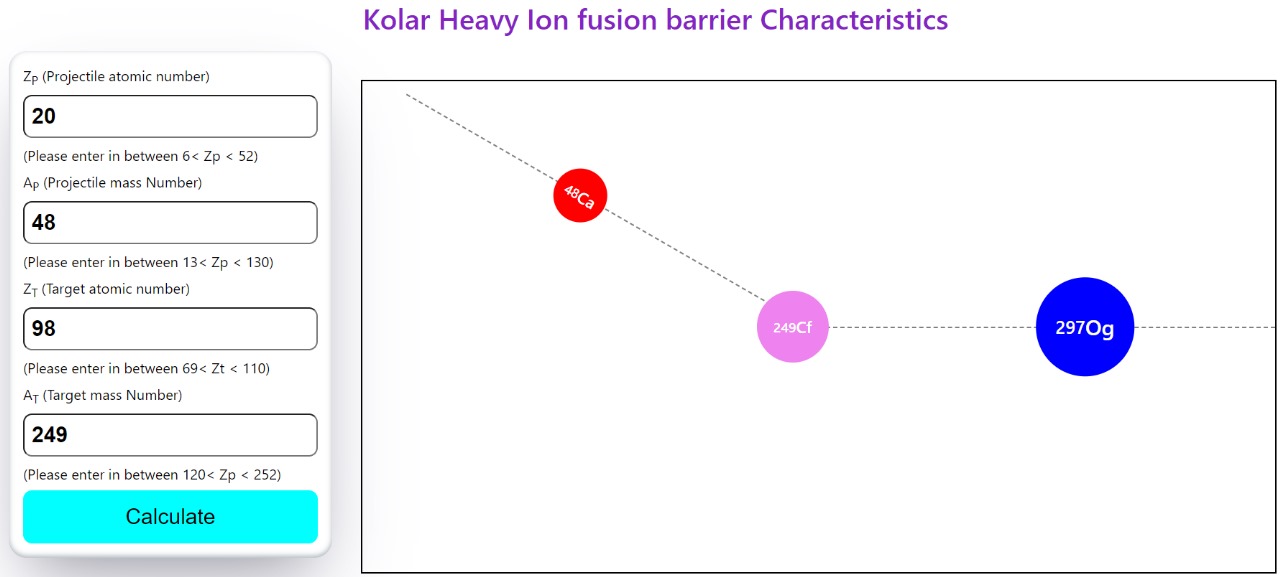}
    \caption{Input page of KHIFC}
    \label{02}
\end{figure}

\begin{figure}[]
    \centering
    \includegraphics[width=\linewidth]{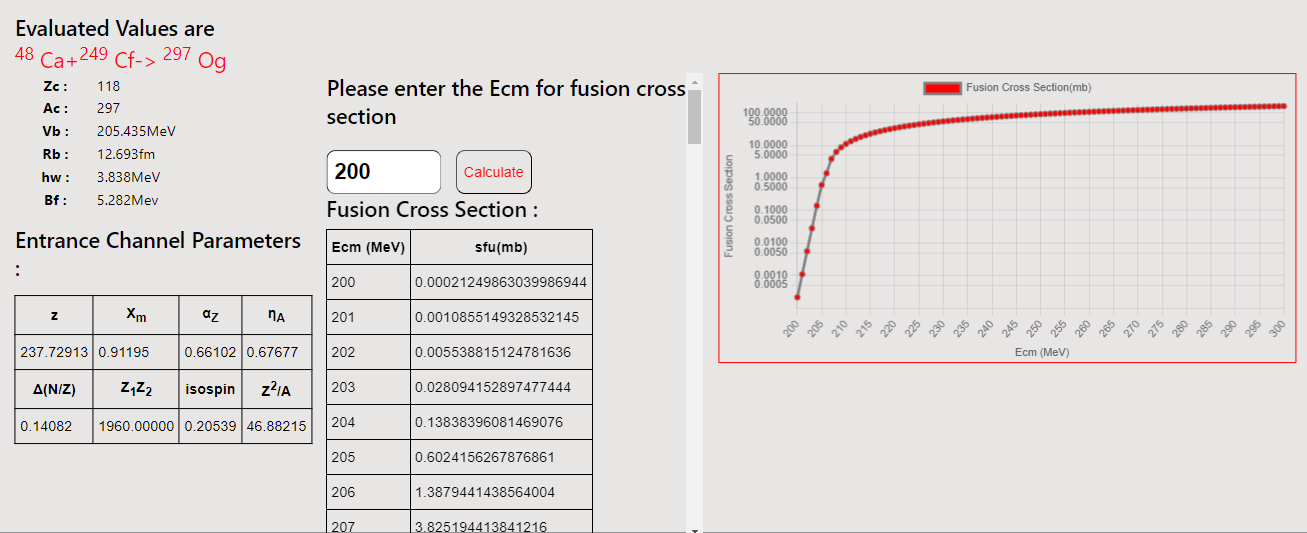}
    \caption{Third page of KHIFC}
    \label{04}
\end{figure}

\indent The constructed program is available in website domain "https://systematics-of-heavy-ion-fusion.vercel.app/". In the first step, the atomic number and mass number of the projectile and target are required to enter in the first page of the KHIFC. The first page of this program is shown in fig. \ref{02}. \\
\begin{figure}[]
    \centering
\includegraphics[width=\linewidth]{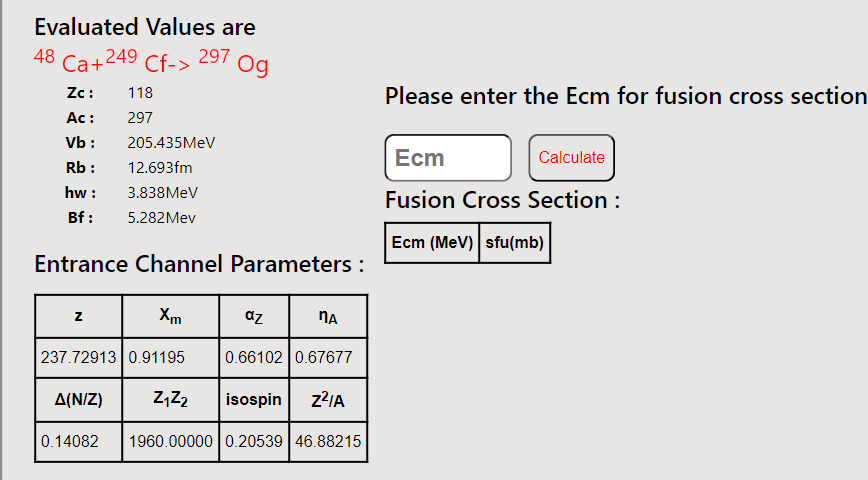}
    \caption{Second page of KHIFC}
    \label{03}
\end{figure}
\indent After entering the projectile and target atomic and mass numbers enables to press the "calculate" button. The apply of "calculate" button goes to second page which is shown in the fig. \ref{03}. This page gives the output of the fusion barrier characteristics such as fusion barrier height ($V_B$ in MeV), fusion barrier radius ($R_B$ in fm), inverted parabola ($\hbar \omega$ in MeV), and fission barrier of the compound nucleus (Bf in MeV). \\
\indent This KHIFC produces the cross-section in the $E_{cm}$ energy range 100 MeV. Input $E_{cm}$ should be greater than $V_B$-20 MeV and it can produce fusion cross-sections for the energy range $E_{cm}$ + 100 MeV. This program also gives the variation of fusion cross-sections with $E_{cm}$ as a graphical representation which is shown in fig. \ref{04}. \\
\begin{figure}[]
    \centering
    \includegraphics[width=\linewidth]{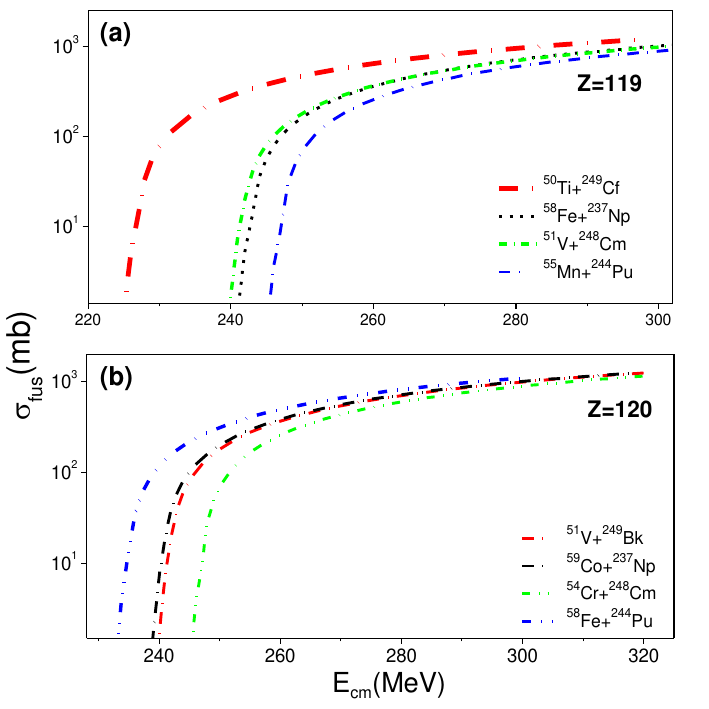}
    \caption{Predicted fusion cross-sections for the synthesis of superheavy element (a) Z=119 and (b) 120 using different fusion cross-sections.}
    \label{comparison}
\end{figure}
\indent Additionally, the fusion cross-sections, calculated using KHIFC, were extended to include superheavy elements with Z=119 and 120. Plots of the fusion cross-sections for the attempted fusion reactions were generated as a function of $E_{cm}$, depicted in Figure \ref{comparison}(a-b). These cross-sections exhibit an increasing trend, reaching peak values at higher center-of-mass energies. Notably, larger fusion cross-sections were observed for the reactions involving $^{50}$Ti+$^{249}$Cf and $^{58}$Fe+$^{244}$Pu, resulting in the formation of Z=119 and Z=120 superheavy elements. These insights not only increase our theoretical knowledge of superheavy element synthesis, but they also give important guidance for future experimental work in this area.\\
\section{Summary}
\indent We have designed simple user  friendly computer program KHIFC  which can be used to study the fusion barrier characteristics of actinide and superheavy nuclei with the simple inputs of atomic and mass numbers of projectile, target, and center of mass-energy.
This program applies to the atomic and mass number range of projectile $3 < Z_P < 51$,  and $6 < A_1 < 127$ respectively.  The target atomic and mass number range varies between $45\le Z_T \le 100$, and $101 \le A_T\le 260$ respectively. The program KHIFC is validated with experiments. Eventhough, this program may have certain deviations less than 2.5$\%$, it can be used to study the fusion characteristics quickly. There is a need for quick calculations before planning to the experiments of heavy element synthesis in such cases this program will be useful.

\bibliography{apssamp}
\bibliographystyle{abbrv}

\end{document}